\documentclass[12pt]{article}

\usepackage{latexsym}
\usepackage{amsfonts,amsmath,amssymb,amsthm}
\usepackage{graphicx}
\usepackage[round]{natbib}
\usepackage{color}

\usepackage[letterpaper,margin=1.5in]{geometry}
\numberwithin{equation}{section}
\newtheorem{thm}{Theorem}

\newtheorem{defi}{Definition}
\newtheorem{lem}[thm]{Lemma}

\newcommand{\act}{\ensuremath{\mathcal A} }
\newcommand{\obs}{\ensuremath{\mathcal O} }
\newcommand{\meas}{\ensuremath{\mathcal M} }
\newcommand{\gspace}{\ensuremath{\mathfrak G} }
\newcommand{\fdg}{\,|\,}
\newcommand{\tr}{\ensuremath{tr}}
\newcommand{\timeo}{\text{\bf O}}
\newcommand{\parents}{\text{pa}}
\newcommand{\children}{\text{ch}}
\DeclareMathOperator*{\ppo}{PpO}
\DeclareMathOperator*{\argmax}{argmax}
\newcommand\cdju[2]{\bigcup_{#1}^{#2}\hspace{-3.9mm\cdot}\hspace{2.2mm}}

\definecolor{lightgrey}{rgb}{0.8,0.8,0.8}

\begin{document}

\thispagestyle{empty}
\title{Structure Learning in Nested Effects Models}
\author{A. Tresch\footnote{corresponding author:
\texttt{tresch@imbei.uni-mainz.de}. Johannes Gutenberg-University
Mainz. Institute for Medical Biometry, Epidemiology and Informatics.
55101 Mainz, Germany} \and F.
Markowetz\footnote{\texttt{florian@genomics.princeton.edu}.
Princeton University. Lewis-Sigler Institute for Integrative
Genomics and Department of Computer Science. Princeton, NJ 08540,
USA}} \maketitle

\begin{abstract}
Nested Effects Models (NEMs) are a class of graphical models
introduced to analyze the results of gene perturbation screens. NEMs
explore noisy subset relations between the high-dimensional outputs
of phenotyping studies, e.g. the effects showing in gene expression
profiles or as morphological features of the perturbed cell.

In this paper we expand the statistical basis of NEMs in four
directions: First, we derive a new formula for the likelihood
function of a NEM, which generalizes previous results for binary
data. Second, we prove model identifiability under mild assumptions.
Third, we show that the new formulation of the likelihood allows to
efficiently traverse model space. Fourth, we incorporate prior
knowledge and an automated variable selection criterion to decrease
the influence of noise in the data.

\vspace{10mm} This manuscript will appear in \emph{Statistical
Applications in Genetics and Molecular Biology} at
\texttt{http://www.bepress.com/sagmb/}
\end{abstract}

\newpage
\section{Introduction}

Functional genomics has a long tradition of inferring the inner
working of a cell through analysis of its response to various
perturbations. There are several perturbation techniques suitable
for large-scale analysis in different organisms. Experiments with
gene knock-outs have been very successful in uncovering gene
function \citep{hughes00functional}, and gene silencing by RNA
interference \citep{fire98potent} allows perturbation screens on a
genome-wide scale.

The changes observed in the cell are called the {\em phenotype} of
the perturbation. In most biological studies, perturbation effects
are measured by single reporters like cell death or growth. Analysis
of these phenotypes can reveal which genes are {\em essential} for
an organism \citep{boutros04genomewide} or for a particular pathway
\citep{gesellchen05rnai}. However, these screens do not reveal {\em
how} the genes contribute to regulatory networks or signalling
pathways.

More details about gene function and interactions are contained in
high-dimensional phenotypes that give a global view of changes in
the cell. High-dimensional phenotypes include gene expression
profiles
\citep{hughes00functional,boutros02sequential,driessche05epistasis},
metabolite concentrations \citep{raamsdonk01functional}, sensitivity
to cytotoxic or cytostatic agents \citep{brown06global}, or
morphological features of the cell \citep{ohya05highdimensional}.
While high-dimensional phenotypic profiles promise a comprehensive
view of the function of genes in a cell, only limited work has been
done so far to adapt statistical and computational methodologies to
the specific needs of large-scale and high-dimensional phenotyping
screens.

\paragraph{Phenotypic profiles offer only indirect information}
A key obstacle to inferring genetic networks from high-dimensional
perturbation screens is that phenotypic profiles generally offer
only indirect information on how genes interact. Cell morphology or
sensitivity to stresses are global features of the cell, which are
hard to relate directly to the genes contributing to them. Gene
expression phenotypes also offer an indirect view of pathway
structure due to the high number of post-transcriptional regulatory
events like protein modifications. For example, when silencing a
kinase we might not be able to observe changes in the activation
states of other proteins involved in the pathway. The only
information we may get is that genes downstream of the pathway show
expression changes. Thus, phenotypic profiles may provide only an
indirect view of information flow and pathway structure in the cell.

\paragraph{Statistical analysis of phenotyping screens}
Previous work focused on clustering phenotypic profiles to find
groups of genes that show similar effects when perturbed. The
rationale is that genes with similar perturbation effects are
expected to be functionally related. The most prominent method used
is average linkage hierarchical clustering
\citep{piano02gene,ohya05highdimensional}. A complementary approach
is ranking genes according to similarity with a query gene
\citep{gunsalus04rnaidb}. In a supervised setting, first steps have
been taken to classify genes into functional groups based on
phenotypic profiles \citep{ohya05highdimensional}. A comprehensive
overview of computational models for the reconstruction of genetic
networks can be found in \citep{markowetz07review}.

A recent approach especially designed to learning from indirect
information and high-dimensional phenotypes are Nested Effects
Models \citep{markowetz05nontranscriptional,markowetz07nested} that
reconstruct features of the internal organization of the cell from
the nested structure of observed perturbation effects. Perturbing
some genes may have an influence on a global process, while
perturbing others affects sub-processes of it. Imagine, for example,
a signaling pathway activating several transcription factors.
Blocking the entire pathway will affect all targets of the
transcription factors, while perturbing a single downstream
transcription factor will only affect its direct targets, which are
a subset of the phenotype obtained by blocking the complete pathway.
NEMs can be seen as a generalization of similarity-based clustering,
which orders (clusters of) genes according to subset relationships
between the sets of phenotypes. So far, a likelihood function has
been derived for NEMs in the case of discretized or binary data
\citep{markowetz05nontranscriptional} and $p$-values of differential
expression \citep{froehlich07estimating}. For model inference, {\em
divide-and-conquer} strategies have been applied to scale up model
search \citep{markowetz07nested,froehlich07large}.

\paragraph{Overview of this paper}
After introducing a generalized version of NEMs in
Section~\ref{section:definition} we expand their statistical basis
in four directions: First, we derive a new formula for the
likelihood function of a NEM that generalizes previous results
(Section~\ref{section:MAPestimation}). Second, we prove model
identifiability under mild assumptions
(Section~\ref{section:identifiability}). Third, we develop efficient
methods of traversing model space
(Section~\ref{section:modelsearch}). And finally, we incorporate
prior knowledge and a variable selection step into model search to
decrease the influence of noise in the data
(Section~\ref{section:extensions}). We show the applicability of the
proposed method in the controlled setting of a simulation scenario
(Section~\ref{section:simulation}) and in an application to an
example in {\em Drosophila} immune response
(Section~\ref{section:application}).

\section{Definition of nested effects
models}\label{section:definition}

The system of components we consider consists of a set \obs of
$n_{\obs}$ observable entities ({\em e.g.} mRNA concentrations), and
a set \act of $n_{\act}$ actions ({\em i.e.} gene perturbations)
applied to the system which are expected to alter the state of some
observable entities. Both \obs and \act consist of binary variables.
An altered state of an observable $s\in \obs$ is denoted by $s=1$,
the basic state is $s=0$. A value of $a=1$, resp. $a=0$, means that
action $a\in \act$ was performed, resp. not performed. Let $D_{as}$,
$(a,s)\in \meas \subseteq \act\times\obs$ be a set of measurements
for observation $s$ after performing action $a$. The set of all
measurements $D=\{D_{as}\fdg (a,s)\in\meas\}$ constitutes the data.
Note that our definition of $\meas$ does not require that {\em all}
$s\in \obs$ are observed for all actions $a\in \act$. Thus, missing
data and the exclusion of failed experiments can directly be
incorporated into all the results that we develop in the following.

\begin{defi}
A (general) {\em effects model} is a binary $n_\act \times n_\obs$
matrix $F$ that determines the state of the observable $s$ when
action $a$ is performed, an entry $0$ indicating no change, $1$
indicating a change.
\end{defi}

Nested effects models are effects models that can be defined in
terms of two graphs or adjacency matrices. The first graph,
$\Gamma$, describes how actions imply each other and the second
graph, $\Theta$, how observables are linked to actions. Let the
actions graph $\Gamma = (\Gamma_{aa'})$ be a graph on the vertices
\act, encoded as an $n_\act \times n_\act$ adjacency matrix with the
convention $\Gamma_{aa}=1$, $a\in\act$. We say that the edge $a\to
a'$ is in $\Gamma$, or for short $a\stackrel{\Gamma}{\to}a'$, if
$\Gamma_{aa'}=1$.

Secondly, we assume that each observation is directly linked to
exactly one action as defined by a function $\theta:\obs \to \act$.
This can synonymously be encoded as an $n_\act \times n_\obs$
adjacency matrix $\Theta = (\Theta_{a s})$, with $\Theta_{a
s}=\delta_{a=\theta(s)}$ for $a\in\act$, $s\in \obs$ (where
$\delta_\cdot$ is the delta function). Write
$a\stackrel{\Theta}{\to}s$ if $\Theta_{as}=1$. By this definition,
$\Theta$ contains only zeros except for a single $1$ in each column.
When describing how observables are linked to actions, we tacitly
switch between the adjacency matrix $\Theta$ and the function
$\theta$ for the sake of notational convenience.

We postulate an effect of an action $a\in \act$ on $s\in \obs$ if
and only if there exists an action $a'\in \act$ such that the edge
from $a$ to $a'$ is in $\Gamma$, and $s$ is directly linked to $a'$
(the edge from $a'$ to $s$ is in $\Theta$). Since each observable is
linked to exactly one action, action $a$ has an effect on $s$ if and
only if $(\Gamma\Theta)_{a s}=1$. This prompts the following
definition:

\begin{defi}
A {\em nested effects model} (NEM) $F$ is an effects model which can
be represented as a product of $\Gamma$ and $\Theta$ as defined
above:
\begin{eqnarray}\label{fdefinition}
F\ =\ \Gamma\,\Theta .
\end{eqnarray}
\end{defi}

\begin{figure}[t]

\includegraphics[width=.23\textwidth]{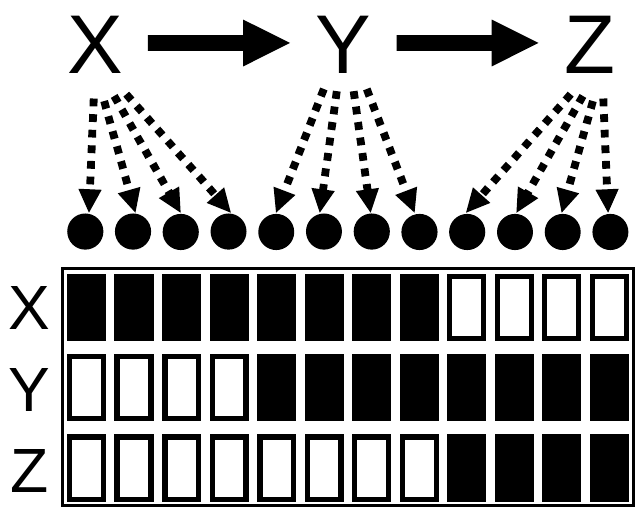}
\hfill
\includegraphics[width=.23\textwidth]{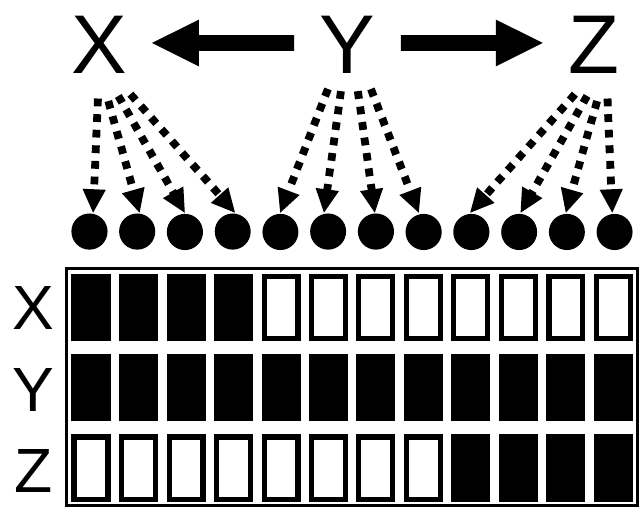}
\hfill
\includegraphics[width=.23\textwidth]{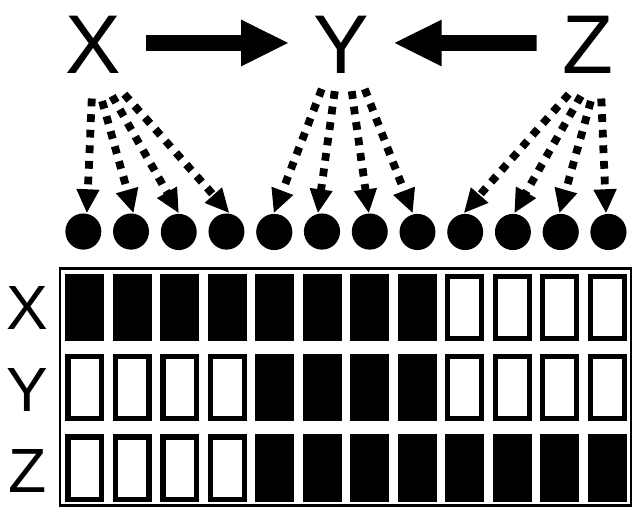}
\hfill
\includegraphics[width=.23\textwidth]{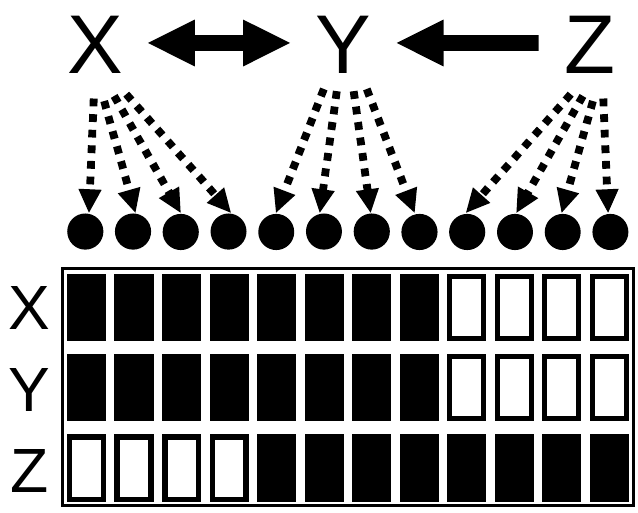}
\caption{NEM examples. Each plot shows a NEM on $\act = \{ X,Y,Z\}$
and twelve observables $\obs$ depicted as dots. The solid arrows
define the adjacency matrix $\Gamma$. Note that cyclic graphs are
allowed, like the right-most example which contains a bi-directional
edge. Each action in $\act$ is connected to observables $s \in \obs$
by dashed arrows encoding $\Theta$ (which is assumed fixed in all
three plots). The lower parts of the plots show the model matrix $F$
in each of these three cases. Each cell in the matrix corresponds to
the state of one observable $s \in \obs$ after performing an action
$a \in \act$. Cells containing observed effects are colored black,
those containing non-effects are white. The objective of structure
learning in NEMs is to recover $\Gamma$ and $\Theta$ from noisy
observations of effect patterns. }\label{NEMexamples}
\end{figure}

The parameters $\Gamma$ and $\Theta$ uniquely determine the model.
We therefore use $P(D\fdg\Gamma,\Theta)$ interchangeably with
$P(D\fdg F)$. Examples of nested effects models are given in
Fig.~\ref{NEMexamples}, showing the graphs $\Gamma$ and $\Theta$ as
well as the resulting effects model $F$.

\paragraph{Transitivity requirement}
Previous definitions of NEMs required the actions graph to be
transitively closed
\citep{markowetz05nontranscriptional,markowetz07nested}, restricting
model space from the space of all graphs to the space of
transitively closed graphs. This is sensible if paths in the actions
graph are interpreted as \emph{causal chains}. For transitively
closed graphs, our model has a property that motivated the name
``nested effects model": The existence of an edge $a \rightarrow a'$
in a transitively closed graph implies that the effects observed for
action $a'$ (i.e. all effects $s$ with $\theta(s)=a'$) are ``nested"
in the observed effects of $a$:
\begin{equation}
a\stackrel{\Gamma}{\to}a' \quad \Leftrightarrow
\quad\{s\in\obs\fdg F_{a's}=1\} \subseteq \{s\in\obs\fdg F_{as}=1\}.
\end{equation}
This correspondence induces an order homomorphism from $\Gamma$ to
the subset lattice of the set observations, which is satisfactory
from a mathematical point of view.

However, admitting only transitively closed graphs as valid models
is a constraint which makes structure learning computationally hard
\citep{markowetz07nested}. Even a small change in the model---like
removing or adding an edge---can make many more changes necessary to
preserve transitivity. The likelihood function will be quite
volatile and the likelihood landscape will not be smooth.

Our calculations here do not rely on the transitivity of the actions
graph. Our definition of NEMs thus extends the one used in previous
studies
\citep{markowetz05nontranscriptional,markowetz07nested,froehlich07estimating,froehlich07large}.

\section{Inference of the actions graph}\label{section:MAPestimation}

\paragraph{Likelihood} Assuming data independence, the likelihood of the model $F$ (with data $D$ fixed) factors into

\begin{eqnarray}\label{likelihood}
P(D|F)\ &=& \prod_{(a,s)\in \meas}\!\! P(D_{as}\fdg s=F_{as})\\&
\propto & \prod_{(a,s)\in\, \obs\times \act}\!\!\!\! P(D_{as}\fdg
s=F_{as})\\
\label{loglikelihood} \text{or\ \ \ \ \ }\log P(D|F)\ &=&
\sum_{(a,s)\in\, \act\times \obs}\!\!\!\! \log\,P(D_{as}\fdg
s=F_{as})\ \ +\ const ,
\end{eqnarray}
if we define $P(s=x|a)=0.5$ for $x\in\{0,1\}$ and $(a,s)\in
(\act\times \obs)\setminus \meas$. The quantity $\log P(D|F)$ can be
expressed in a convenient form: For an observable $s\in \obs$ and a
perturbation $a\in\act$, let the log likelihood ratio $R_{sa} = \log
\frac{P(D_{as}\fdg s=1)}{P(D_{as}\fdg s=0)}$ be known, and
$R=(R_{sa})$ be the $\obs \times \act$ matrix of ratios. If we let
$N$ be the null matrix, i.e. the model predicting no effects at all,
then
\begin{eqnarray}\label{trace1}
\log P(D|F)-\log P(D|N) &\underset{(\ref{loglikelihood})}{=}& \sum_{(a,s)\in\, \act\times \obs} log\,\frac{P(D_{as}\fdg s=F_{a s})}{P(D_{as}\fdg s=0)} \nonumber\\
      &=& \sum_{(a,s)\in\, \act\times \obs}  \left\{\! \begin{tabular}{cl} $R_{s a}$\!\! & if $F_{a s}=1$ \\ $0$ & if $F_{a s}=0$
      \end{tabular}\right. \nonumber\\
      & = & \sum_{a\in \act} \sum_{s\in\,\obs} F_{a s} R_{s a} \nonumber\\
     & = & \sum_{a\in \act} (F R)_{a a} \ = \ \tr\,(F R)\ \ ,
\end{eqnarray}
with ``\tr " denoting the trace function of a quadratic matrix. This
derivation of the likelihood applies to both general effects models
and nested effects models. In particular, in nested effects models
Eq.~(\ref{fdefinition}) allows to represent the likelihood as
\begin{eqnarray}\label{trace2}
\log P(D|\Gamma,\theta)\ =\ \tr (\Gamma \Theta R)\ +\ const \ .
\end{eqnarray}
The likelihood function depends on the
data only via the likelihood ratios in $R$. This makes our approach
very flexible: our method can handle as input data binary values,
$p$-values, or any other arbitrary statistic as long as it can be
converted to a likelihood ratio. Section~\ref{rmatrix} contains an
outline of how this quantity can be estimated in a practical
application to gene expression microarray data.

\paragraph{Posterior}
We aim at maximizing the posterior of $\Gamma$ and $\Theta$,
\begin{eqnarray}\label{posterior}
P(\Gamma,\Theta \fdg D) &=& \frac{P(D\fdg \Gamma,\Theta)\cdot
P(\Gamma,\Theta)}{P(D)} \nonumber\\ & \propto & P(D\fdg
\Gamma,\Theta)\cdot P(\Gamma)\cdot P(\Theta),
\end{eqnarray}
where we assume that the parameters $\Gamma$ and $\Theta$ are
independent and follow prior distributions $P(\Gamma)$ amd
$P(\Theta)$, which are not necessarily uniform.

Let $Q=(Q_{sa})$ be an $n_\obs \times n_\act$ matrix with entries
$Q_{sa}=P(\theta(s)=a)$. We assume that the prior links each
observation independently to an action, i.e.
\begin{eqnarray}\label{priortheta} P(\Theta) =
\prod_{s\in\obs}P(\theta(s)=a_s)\ \ \ \text{or}\ \ \ \log P(\theta)
= \sum_{s\in\obs} Q_{sa_s}
\end{eqnarray}
where $a_s$ is the particular value of $\theta(s)$ in $\Theta$.
Consider the data $D$ fixed and write $L(\Gamma,\theta) = \log P(D
\fdg \Gamma,\theta )$ for the log-likelihood of the data, given the
model. Then the posterior of the model $(\Gamma, \theta)$ becomes
\begin{eqnarray}
\log P(\Gamma,\theta\fdg D) &\underset{(\ref{posterior})}{=}& L( \Gamma,\theta ) \ +\  \log P(\Gamma) \ +\ \log P(\theta)\ +\ const
\end{eqnarray}

The task is to find the MAP estimate for $P(\Gamma,\Theta\fdg D)$,
\begin{eqnarray}\label{mapestimate}
(\hat{\Gamma },\hat{\theta}) &=& \underset{\Gamma,\theta }{\argmax}
\big(\, L(\Gamma,\theta) + \log P(\Gamma) + \log P(\theta) \,\big)
\end{eqnarray}
We are particularly interested in finding the optimal actions graph $\hat{\Gamma}$. Writing
\begin{eqnarray}\label{thetaopt}
\theta_{\Gamma} & = & \underset{\theta}{\argmax}\, \big(
L(\Gamma,\theta) + \log P(\theta) \big)\ \text{,}\\L(\Gamma) & = &
L(\Gamma,\theta_{\Gamma}) = \underset{\theta}{\text{max}}\,
L(\Gamma,\theta) \ ,
\end{eqnarray}
this corresponds to finding
\begin{eqnarray}
\hat{\Gamma } &=& \underset{\Gamma }{\argmax} \big(\,
\underset{\theta }{\text{max}} \big( L(\Gamma ,\theta ) + \log
P(\theta)\big) +\log P(\Gamma ) \,\big)\nonumber\\
 & = & \underset{\Gamma }{\argmax} \big(\, L(\Gamma ) + \log P(\Gamma ) \,\big)
\end{eqnarray}

\section{Model identifiability}\label{section:identifiability}

We present theorems showing that the maximum likelihood estimator
recovers the true structure of the actions graph for sufficiently
``good" data. All proofs are given in the appendix.

\begin{defi}Let some data be observed from the underlying true effects
model $F$. Let $R$ be the ratio matrix which has been derived from
the data. We say that the data is {\em consistent with} $F$ if the
ratio matrix $R$ has a positive entry $R_{sa}$ (= favors an effect)
whenever $F$ has a positive entry $F_{as}$ (= predicts an effect) at
the corresponding position.
\end{defi}

\begin{thm}\label{consistency} If the data is consistent with the effects model $F$, then the maximum likelihood estimate of (\ref{trace1}) equals $F$,
\begin{eqnarray}
 F =\ \underset{G}{\argmax}\ P(D|G)\ \underset{(\ref{trace1})}{=}\  \underset{G}{\argmax}\ \tr(GR)\ .
 \end{eqnarray}\hfill$\Box$
\end{thm}

In the light of this theorem it is interesting to find out to what
extent the actions graph $\Gamma$ and the assignment $\Theta$ are
controlled by the nested effects model $F=\Gamma\Theta$. The
complete answer is given in Theorem \ref{identifiability}. We
precede it by a definition and a lemma.

\begin{figure}[t]
\begin{center}
\includegraphics[width=.23\textwidth]{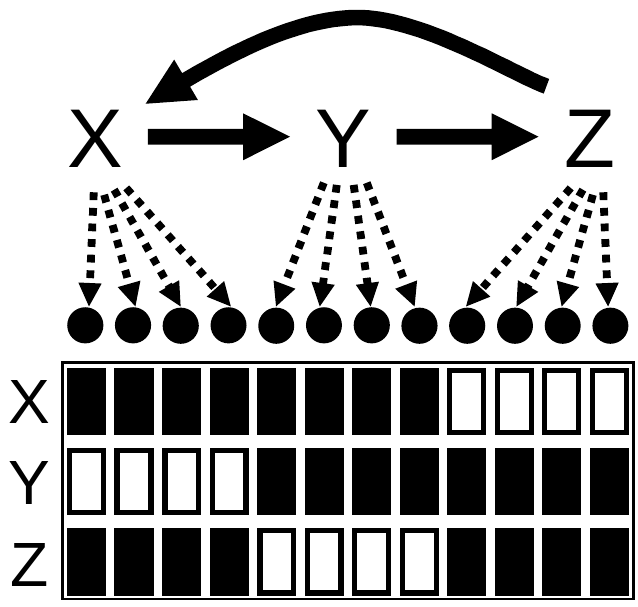}
\hspace{1cm}
\includegraphics[width=.23\textwidth]{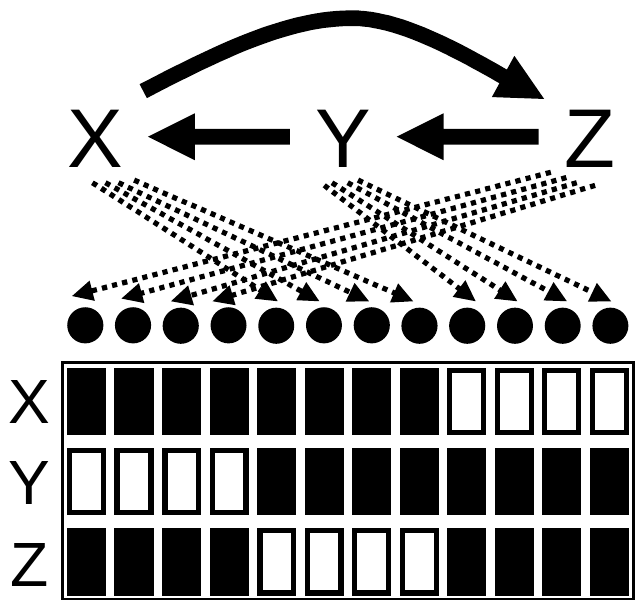}
\end{center}
\caption{\label{fig:reversal}A basic example of a reversal. Both
plots show a NEM on $\act = \{ X,Y,Z\}$ and twelve observables
$\obs$ depicted as dots. The actions graphs $\Gamma$ and ${\Gamma}'$
are both cyclic, but the two models differ in the direction of the
cycle (clock-wise on the right, against the clock on the left).
However, the two model matrices $F$ and $F'$ are identical, since
the change from $\Gamma$ to ${\Gamma}'$ can be compensated by
simultaneously changing the assignment between actions and
observables (indicated by the dashed arrows).}
\end{figure}

\begin{defi}\label{reversals} Let $F$ be a nested effects model
parametrized by $(\Gamma,\Theta)$. Let $a_1\to a_{2}\to ...\to
a_n\to a_1$ be a cycle in $\Gamma$, let $\pi$ be the circular
permutation $\pi=(a_1\,a_2...a_n)$. Let $e_b$ denote the $b$-th unit column vector of length
$n_{\obs}$, and let $S=\sum_{b\in \act} e_b
e^T_{\pi(b)}$ be the permutation matrix corresponding to $\pi$.  We
say that $(\Gamma',\Theta')=(\Gamma S^{-1},S\Theta)$ is a {\em reversal} of $(\Gamma,\Theta)$ induced by $\pi$ (see
Fig.~\ref{fig:reversal} for an example). Two
reversals are called {\em disjoint} if they are induced by disjoint
cyclic permutations (i.e. each action is fixed by at least one of
the permutations).
\end{defi}

\begin{lem}\label{lemma:reversal}
Let $(\Gamma',\Theta')$ be a reversal of $(\Gamma,\Theta)$ induced by the permutation $\pi = (a_1,a_2,...,a_n)$. Then $(\Gamma',\Theta')$ is a valid parametrization
of $F=\Gamma\Theta$.
\end{lem}

Lemma \ref{lemma:reversal} states that the two parametrizations
$(\Gamma,\Theta)$ and $(\Gamma',\Theta')$ define the same nested
effects model, thus we call them {\em observationally equivalent}.

For an action $a\in \act$, the {\em parents} of $a$ are the actions $b\in \act$ such that $b\stackrel{\Gamma}{\to}a$. If
 two distinct actions $a,b\in\act$ have the same parents, then they are clearly indistinguishable by any
kind of interventional measurement. This is a general limitation,
not only a limitation of our model. We propose collapsing these two
actions in such a case \citep{markowetz07nested}. We exclude
indistinguishable actions from our considerations and state:

\begin{thm}\label{identifiability}
Let $(\Gamma,\Theta)$ and $(\Gamma',\Theta')$ be the parameters of
two nested effects models. Assume that no two distinct actions
$a,b\in\act$ have the same parents in $\Gamma$ or in $\Gamma'$. Then
$(\Gamma,\Theta)$ and $(\Gamma',\Theta')$ are observationally
equivalent if and only if the tuples can be converted one into
another by a sequence of disjoint reversals.\hfill$\Box$
\end{thm}

Taken together, Theorems \ref{consistency} and \ref{identifiability}
state that under mild conditions, not only the true nested effects
model $F$ is identifiable for "sufficiently good" data (which in
practical cases means for a sufficiently high number of replicate
measurements), but also the underlying actions graph $\Gamma$ and
the assignment $\theta$ are unique up to reversals.

\section{Actions graph search}\label{section:modelsearch}

The space \gspace of actions graphs is huge, it contains
$2^{n_{\act}(n_{\act}-1)}$ elements (recall that the diagonal
entries of the adjacency matrix equal $1$\,). In order to search
\gspace efficiently, we need a fast method for the evaluation of
(\ref{posterior}). The key observation is that small changes in
$\Gamma$ require only small changes in the optimal assignment
$\theta$, and these can be calculated very fast.

\paragraph{Elementary moves}
Let $\Gamma$ and $\Gamma'$ be {\em neighbours} in \gspace, i.e.
$\Gamma$ and $\Gamma'$ differ exactly in one edge, from action $j$
to action $k$ say. An {\em elementary move} is defined as the
insertion or the removal of one edge in an actions graph (note that
if we chose the search space \gspace to be the transitively closed
graphs or the acyclic graphs, these moves would not be well
defined). Let $e_j$ denote the $j$-th unit column vector of length
$n_{\obs}$, and let $g_k$ be the $k$-th unit column vector of length
$n_{\act}$. By $x^T$ denote the transpose of a vector or a matrix
$x$. Define $E_{jk}=e_je_k^T$ as the matrix containing only zero
entries except a $1$ in row $j$, column $k$. Then
\begin{eqnarray}\label{onestep}
\Gamma' = \Gamma + \epsilon_{jk}E_{jk}\ ,\ \  \text{where}\ \ \
\epsilon_{jk}=1-2\Gamma\!\!_{jk} .
\end{eqnarray}

\paragraph{Optimization of the effects graph}\label{firstbracket}
Let $G_{js}=e_jg_s^T$, $j\in\act$, $s\in \obs$. Then
\begin{eqnarray}\label{traceformula}
L(\Gamma,\theta)-L(f_0) &\underset{(\ref{trace2})}{=}&  \tr(\Gamma\Theta R)\nonumber  \\ &=& \tr (\Gamma (\sum_{s\in\obs} G_{\theta(s),s}) R ) = \sum_{s\in\obs} \tr(\Gamma  G_{\theta(s),s} R )  \nonumber\\
 &=& \sum_{s\in\obs} \tr(\Gamma  e_{\theta(s)}\cdot g_s^T R ) = \sum_{s\in\obs} \tr (\Gamma_{\cdot \theta(s)} R_{s\cdot}) = \sum_{s\in\obs} \tr ( R_{s\cdot} \Gamma_{\cdot \theta(s)} )    \nonumber\\
\label{updateeffects}
 &=& \sum_{s\in\obs}  R_{s\cdot} \Gamma_{\cdot \theta(s)} = \sum_{s\in\obs}  (R \Gamma)_{s \theta(s)}
\end{eqnarray}
It follows from the equations (\ref{priortheta}),(\ref{thetaopt}) and (\ref{updateeffects}) that maximizing $\theta$ with respect to $L(\Gamma,\theta)$ can be done pointwise, i.e.
\begin{eqnarray} \label{thetamax}
\theta_{\Gamma}(s) = \underset{a\in\act}{\argmax}\big( (R \Gamma)_{s
a} + Q_{sa} \big)\ ,\ \ s\in\obs .
\end{eqnarray}
For each $s\in \obs$, step (\ref{thetamax}) takes $\timeo(n_{\obs})$
time, provided that the matrix $R\Gamma$ is given. It is therefore
necessary to keep track of this matrix whenever $\Gamma$ is changed
into a $\Gamma'$. But $R\Gamma' = R\Gamma + \epsilon_{jk}RE_{jk}$ is
obtained from $R\Gamma$ simply by adding $\epsilon_{jk}R.j$ to the
$k$-th column of $R\Gamma$, so this process takes only
$\timeo(n_{\act})$ time. The complete evaluation of $\theta$
according to (\ref{thetamax}) takes $\timeo(n_{\act} n_{\obs})$
time. However we can exploit the fact that (in expectation) hardly
any of the observable effects has to be reassigned. For the moment,
fix $s\in\obs$ and consider the vector $v=(R\Gamma)_{s\cdot}+
Q_{s\cdot}$ and let
\begin{eqnarray}
 w &=& (R\Gamma')_{s\cdot} + Q_{s\cdot} = \left( R(\Gamma + \epsilon_{jk}E_{jk}) \right)_{s\cdot} + Q_{s\cdot} \nonumber\\ &=& \left( (R\Gamma)_{s\cdot} + \epsilon_{jk}(R e_je_k^T \right)_{s\cdot}) + Q_{s\cdot}\nonumber \\ &=&  v +  \epsilon_{jk}(e_s^T R e_j)e_k^T  = v + \epsilon_{jk}R_{sj} e_k^T \ .\label{technicalvector}
\end{eqnarray}
By (\ref{thetamax}), $t=\theta_{\Gamma}(s) = \argmax_{a \in \act} v_a$ and $\theta_{\Gamma'}(s) = \argmax_{a \in \act} w_a$.  The vector $v$ differs from $w$ at most in its $k$-th entry. The following cases can occur:
\begin{eqnarray} \label{fastupdate}
\theta_{\Gamma'}(s) &=& \left\{
\begin{tabular}{cl}
  $t$ & \text{if $t\neq k$ , $w_k\leq v_t$} \\
  $k$ & \text{if $t\neq k$ , $w_k > v_t$} \\
  $k$ & \text{if $t = k$ , $w_k\geq v_k$} \\
  $\underset{a \in \act}{\argmax }\, w_a$ & \text{if $t = k$ , $w_k < v_k$}
\end{tabular} \right.
\nonumber \\ &=& \! \left\{
\begin{tabular}{cl}
  $t$ & \text{if $t\neq k$ , $v_k+\epsilon_{jk}R_{sj}\leq v_t$} \\
  $k$ & \text{if $t\neq k$ , $v_k+\epsilon_{jk}R_{sj} > v_t$} \\
  $k$ & \text{if $t = k$ , $\epsilon_{jk}R_{sj}\geq 0$} \\
  $\underset{a \in \act}{\argmax }\, w_a$ & \text{if $t = k$ , $\epsilon_{jk}R_{sj} < 0$}
\end{tabular} \right.
\end{eqnarray}
Given the matrix $R\Gamma$, the first three cases in
(\ref{fastupdate}) can be calculated in constant time. The fourth
case requires $\timeo(n_{\act})$ time. The elementary moves choose
every edge $j\to k$ with the same frequency, so the expected
relative frequency for which the case $t=k$ occurs is
$\frac{1}{n_{\act}}$. Therefore, the expected running time for
(\ref{fastupdate}) is at most $\timeo (
\frac{n_{\act}-1}{n_{\act}}\cdot 1 + \frac{1}{n_{\act}}\cdot
n_{\act}) = \timeo(1)$. We have to do this step for all $s\in \obs$,
so the calculation of the function $\theta$ can be done in expected
$\timeo(n_{\obs})$ time. What remains to do is to update the matrix
$R\Gamma$ to \[R\Gamma'=R\Gamma + \epsilon{jk}RE_{jk} = R\Gamma +
\epsilon_{jk}R_{\cdot j}e_k^T \ .\] This only affects the $k$-th
column of $R\Gamma$, to which we add the vector
$\epsilon_{jk}R_{\cdot j}$. The time consumption of this step is
$\timeo(n_{\obs})$.

\paragraph{Gray code enumeration of actions graphs}
Actions graphs are treated as binary vectors of length
$n_{\act}^2-n_{\act}$ (the diagonal is fixed), and they are
enumerated without redundances using a gray code \citep{knuthvol4}.
Each enumeration step alters exactly one edge of the predecessor
graph, so we can take advantage of our fast update algorithm. It
allows the exhaustive search of the actions graph space \gspace for
$n\leq 5$ (computation time on a 1GHz computer: a few seconds for
$n=4$ actions, approx. 10mins for $n=5$ actions).

\section{Extensions}\label{section:extensions}

In this section, we adapt the raw nested effects model to make it
more applicable to real-life data sets. We discuss methods to
incorporate prior knowledge on parts of an action graph and to
decrease measurement noise by feature selection and regularization.

\subsection{Rigid actions graph prior}\label{section:double}

In many practical applications, parts of the true actions graph
structure is already known, and only a fraction of edges has to be
estimated from the data. Taking advantage of this, we introduce a
rigid prior on the actions graph. An edge can be declared as known
present, known absent, or unknown. Exhaustive search is then
performed only on those edges whose presence is unknown.

This permits a novel way of joining new components to a well known
signaling network: Given measurements of a known actions graph and
an additional action node $a$, declare only edges starting or ending
in $a$ as unknown. The reconstruction procedure will then find the
position of $a$ within the already established network. We show the
feasibility of this procedure in the simulations in section
\ref{simulation:rigidprior}.

\subsection{Feature selection and regularization}\label{section:featureselection}

In high-dimensional phenotypic readouts, we may encounter a
situation in which a considerable part of all observables does not
react to any intervention at all. The occurrence of many false
positive effects is an inevitable consequence. Therefore, it is
essential to only include responsive observables into the model and
discard the rest.

\paragraph{The null action}
Our model offers an elegant way of doing feature selection: Extend
the adjacency matrix $\Gamma$ of the actions graph by one null
column, which can be interpreted as an action that does not affect
the observations assigned to it (we call it the {\em null action} in
contrast to the regular actions in \act). The optimization procedure
in Section \ref{section:modelsearch} then assigns a gene to the null
action if considering the gene a general non-responder is beneficial
to the posterior.

This method has two advantages: It does hardly cost any extra
computation time, and the number of responsive genes does not have
to be fixed in advance. For example a best fitting graph structure
might recruit many weakly responsive genes, whereas in other
situations it might receive less numerous but strong support by only
a few genes.

\paragraph{Regularization}
We complement the null action with a noise reducing regularization
step. A straightforward way is to subtract a (non-negative) constant
$\delta$ from each entry in the ratio matrix $R$. This amounts to a
priori favoring non-effects, since
\begin{eqnarray}
    R_{sa} -\delta = \log \frac{P(D_{as}\fdg s=1)\phantom{\cdot \exp(\delta)}}{P(D_{as}\fdg s=0)\cdot
    \exp(\delta)}.
\end{eqnarray}
Suppose that all values $R_{sa}-\delta$, $a\in\act$ in some row of
$R$ are negative. Then any assignment of the observable $s$ to a
regular action will decrease the posterior. Thus, in any model, $s$
will be optimally assigned to the null action. It is therefore
time-saving to directly exclude this effect before entering the
reconstruction algorithm.

\paragraph{The $\ppo$-score}
We propose a simple heuristic for the optimal choice of $\delta$
which works well in practice. Let $\obs(\delta)$ be the effects that
are still included into the reconstruction step after the
regularization by $\delta$ has been applied. The larger $\delta$,
the smaller $\obs(\delta)$, and $\delta=0$ corresponds to no
pre-selection at all. Let
$(\hat{\Gamma_{\delta}},\hat{\Theta_{\delta}})$ be the maximum a
posteriori estimate derived from the ($\delta$-)regularized ratio
matrix $R$. Define the {\em posterior per observable} score
by\begin{eqnarray}
     \ppo(\delta) =
     \frac{L(\hat{\Gamma}_{\delta},\hat{\Theta}_{\delta})}{|\obs(\delta)|}.
\end{eqnarray}
The $\ppo$ measures the average contribution of each effect in
$\obs(\delta)$ to the log posterior value of the best scoring model.
Select  the optimal value of $\delta$ as
\begin{equation}\hat{\delta} = \argmax_\delta \ppo(\delta).\end{equation}
Figure \ref{deltachoice} shows a typical $\ppo$ curve, which
compares models with varying degrees of regularization for the
\emph{Drosophila} data set we will describe in detail in section
\ref{section:application}.

\section{Simulation results}\label{section:simulation}

\subsection{Robustness of the actions graph reconstruction}\label{robustness}
Section \ref{section:identifiability} proved the identifiability of
nested effects models under the assumption of consistent data. Here
we investigate the robustness of NEMs against measurement errors,
i.e. variability in the ratio matrix.

Given a ``true" NEM and a noise level $\alpha$, we calculate a
consistent ratio matrix containing the entry $0.5$ (resp. $-0.5$)
whenever the model predicts an effect (resp. no effect). Then, we
add independent, normally $\mathcal N (0,\alpha^2)$-distributed
noise to each entry of the ratio matrix.

An exhaustive search on the actions graph space produces a
distribution of posterior scores as well as the highest scoring NEM.
From this distribution, we compute the rank of the score of the
original NEM among all scores as well as the number of true positive
and false positive edges in the highest scoring NEM. Results are
averaged over 100 randomly sampled NEMs for each level of noise.

\begin{figure}[t]
\begin{center}
\includegraphics[width=.32\textwidth]{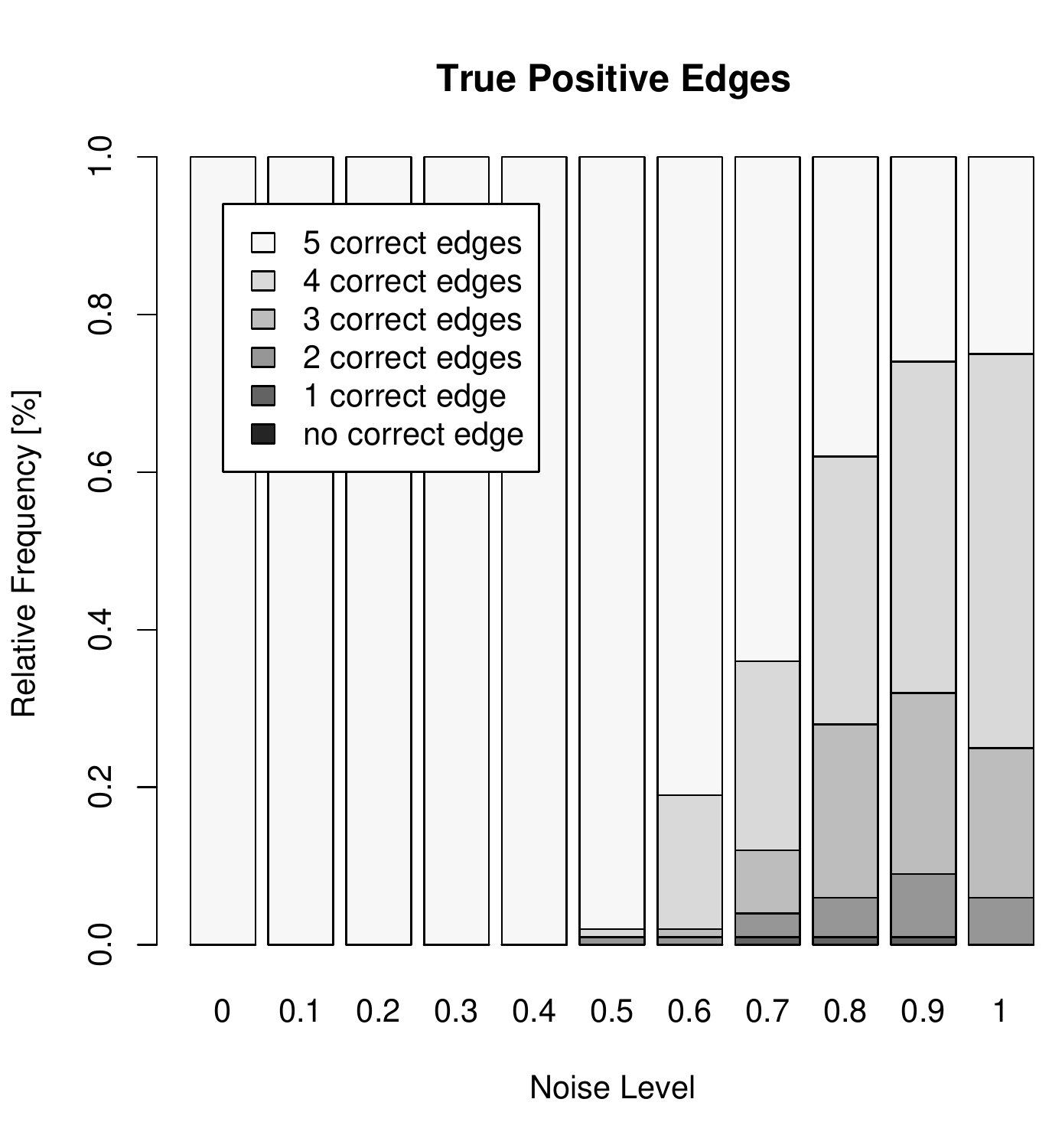}
\hfill
\includegraphics[width=.32\textwidth]{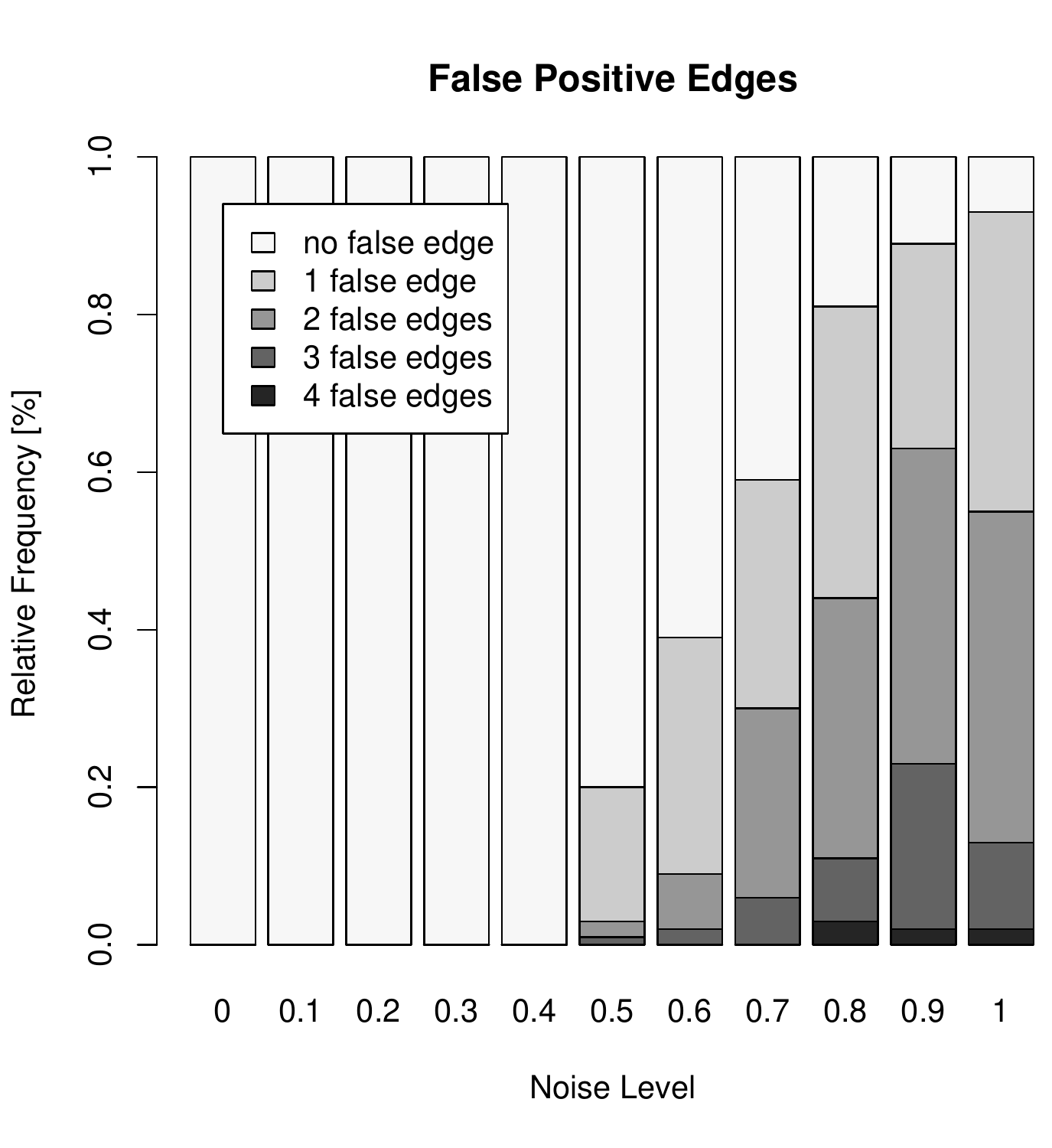}
\hfill
\includegraphics[width=.32\textwidth]{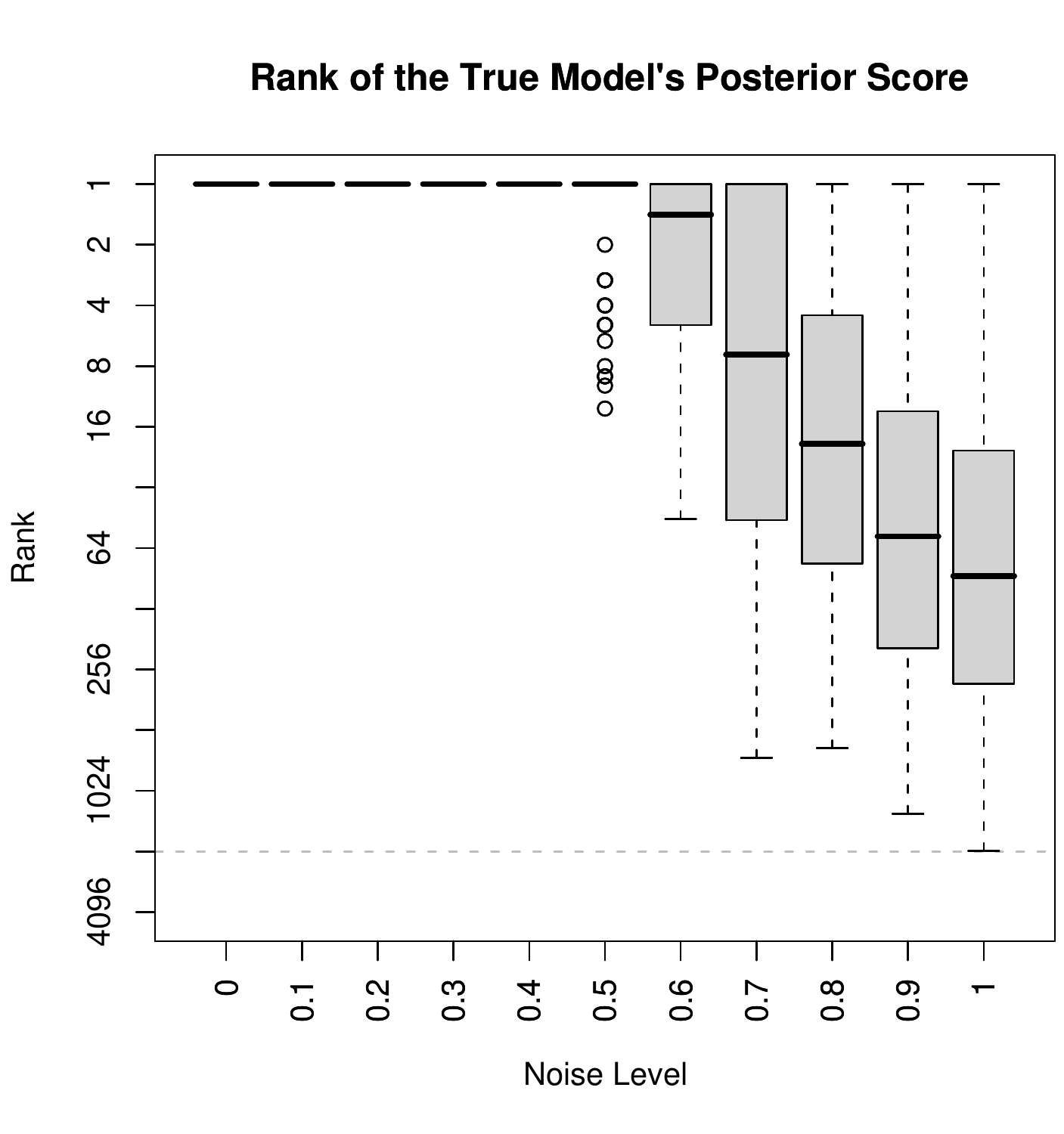}
\end{center}
\caption{Reliability of the actions graph reconstruction in the
presence of noise. The noise level was varied from 0 to 1 in steps
of 0.1. For each level of noise, the respective plot shows the
statistics over 100 sample NEMs containing 5 edges. \textbf{Left:}
The number of true positive edges in the highest scoring NEM.
\textbf{Middle:} The number of false positive edges in the highest
scoring NEM. \textbf{Right:} The (distribution of the) rank of the
posterior score of the true model among all posterior scores. }
\label{increasingnoise}
\end{figure}

The results displayed in Figure \ref{increasingnoise} show the
reliability of reconstructing the actions graph under increasing
noise. Only at noise levels above $0.5$ does our model start to miss
edges (left plot) or to include spurious edges (middle plot) and the
correct graph may then no longer be the highest scoring (right
plot). For noise levels below $0.5$ we achieve perfect
reconstruction in all simulation runs.

\subsection{Utility of prior knowledge}\label{simulation:rigidprior}
We test the impact of prior knowledge on the quality of the actions
graph reconstruction in two ways. First, a fixed ``true" model
consisting of $4$ actions, $50$ observables and $5$ edges is
constructed, and a noisy ratio matrix is generated from it. The
noise level is set to $\alpha = 0.7$.

Starting from this matrix, a series of exhaustive searches is
carried out. Each time, a prior is generated that either fixes a
number of truly present edges as present, or which specifies a
number of truly absent edges as absent. The quality of
reconstruction is assessed in terms of sensitivity and specificity
(regarding only those edges that were not known a priori).  Since
the quality of reconstruction heavily depends on the true actions
graph topology, we average the results over 100 sample runs of this
procedure.

\begin{figure}[t]
\begin{center}
\includegraphics[width=.32\textwidth]{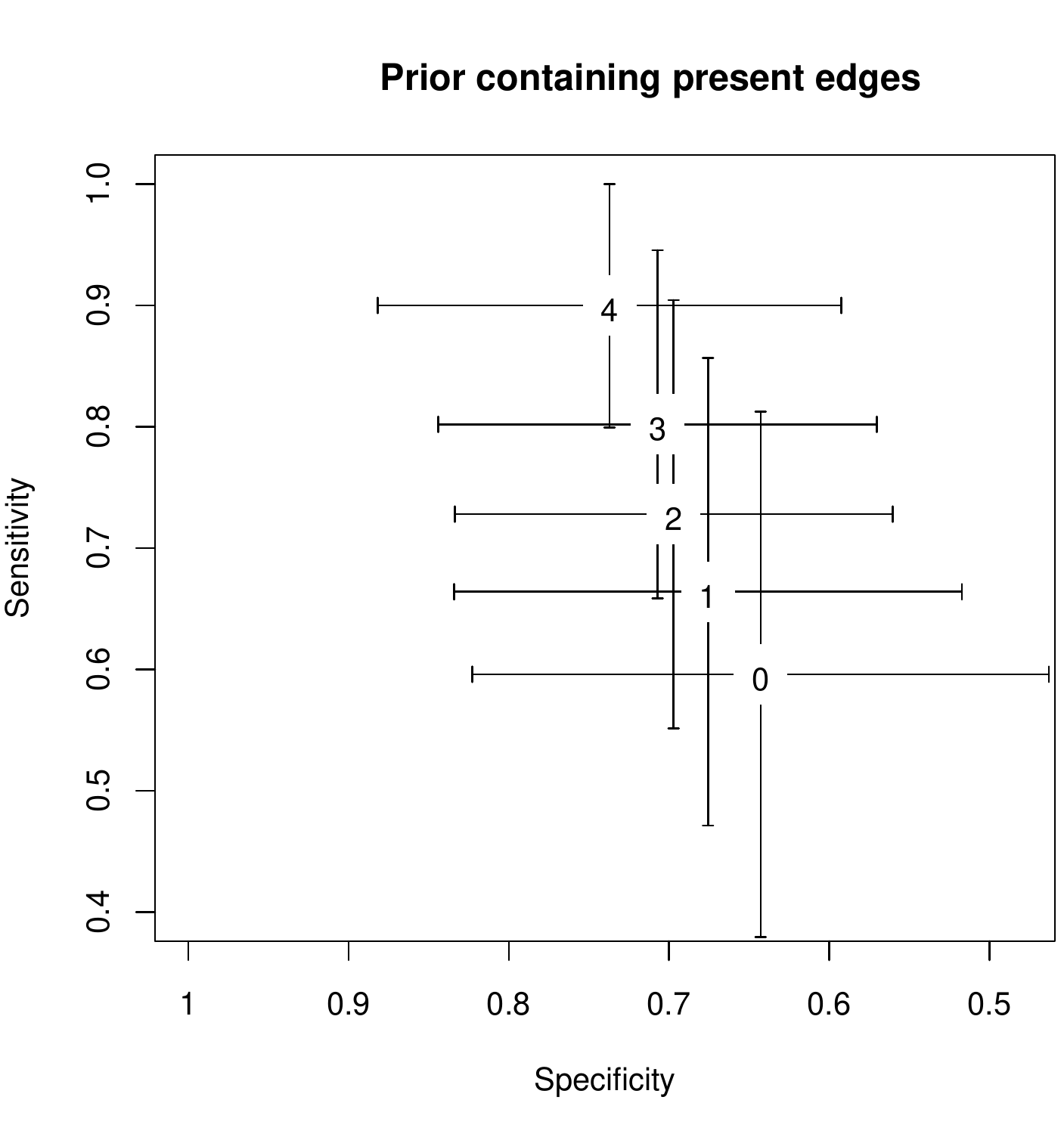}
\hfill
\includegraphics[width=.32\textwidth]{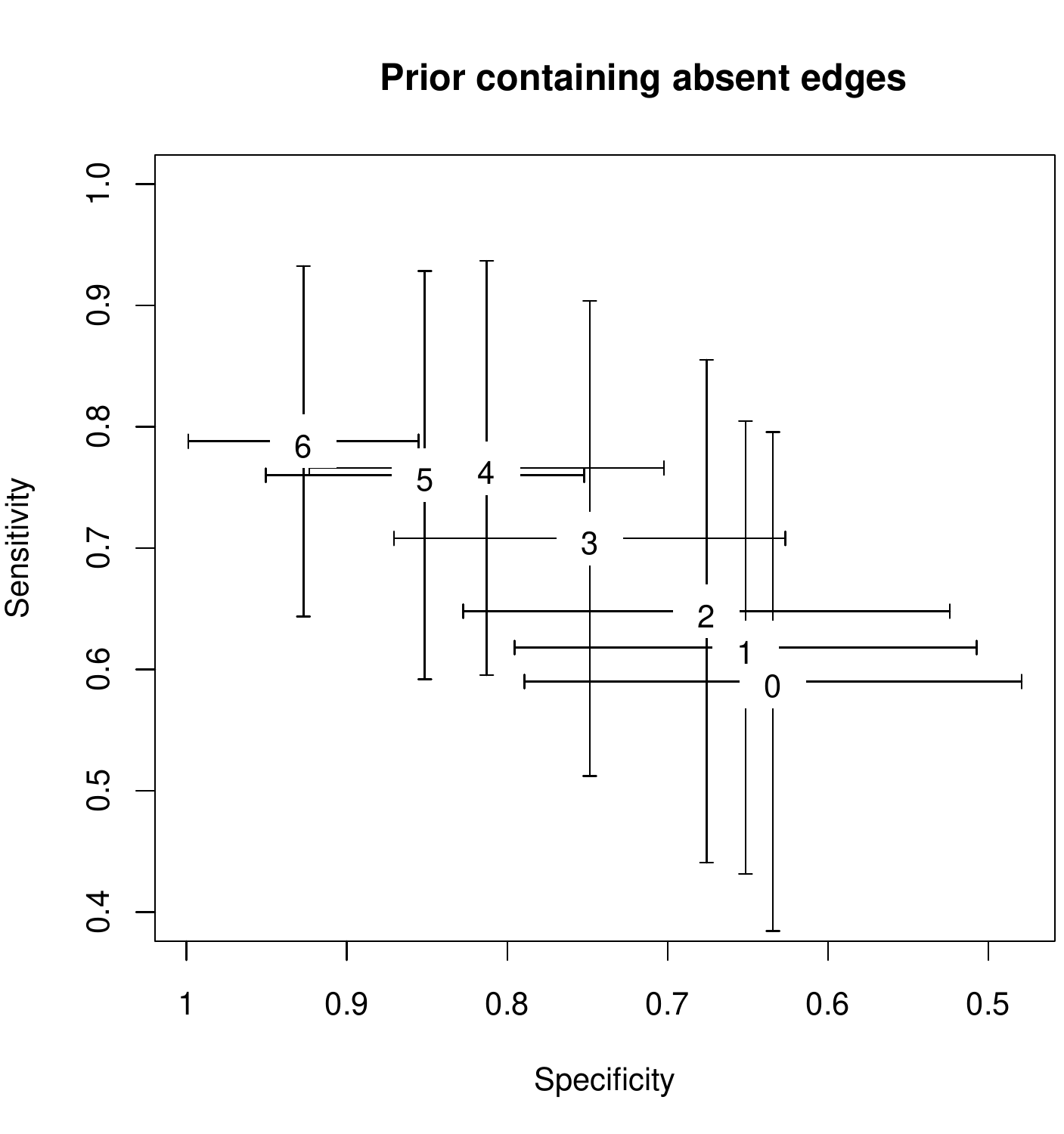}
\hfill
\includegraphics[width=.32\textwidth]{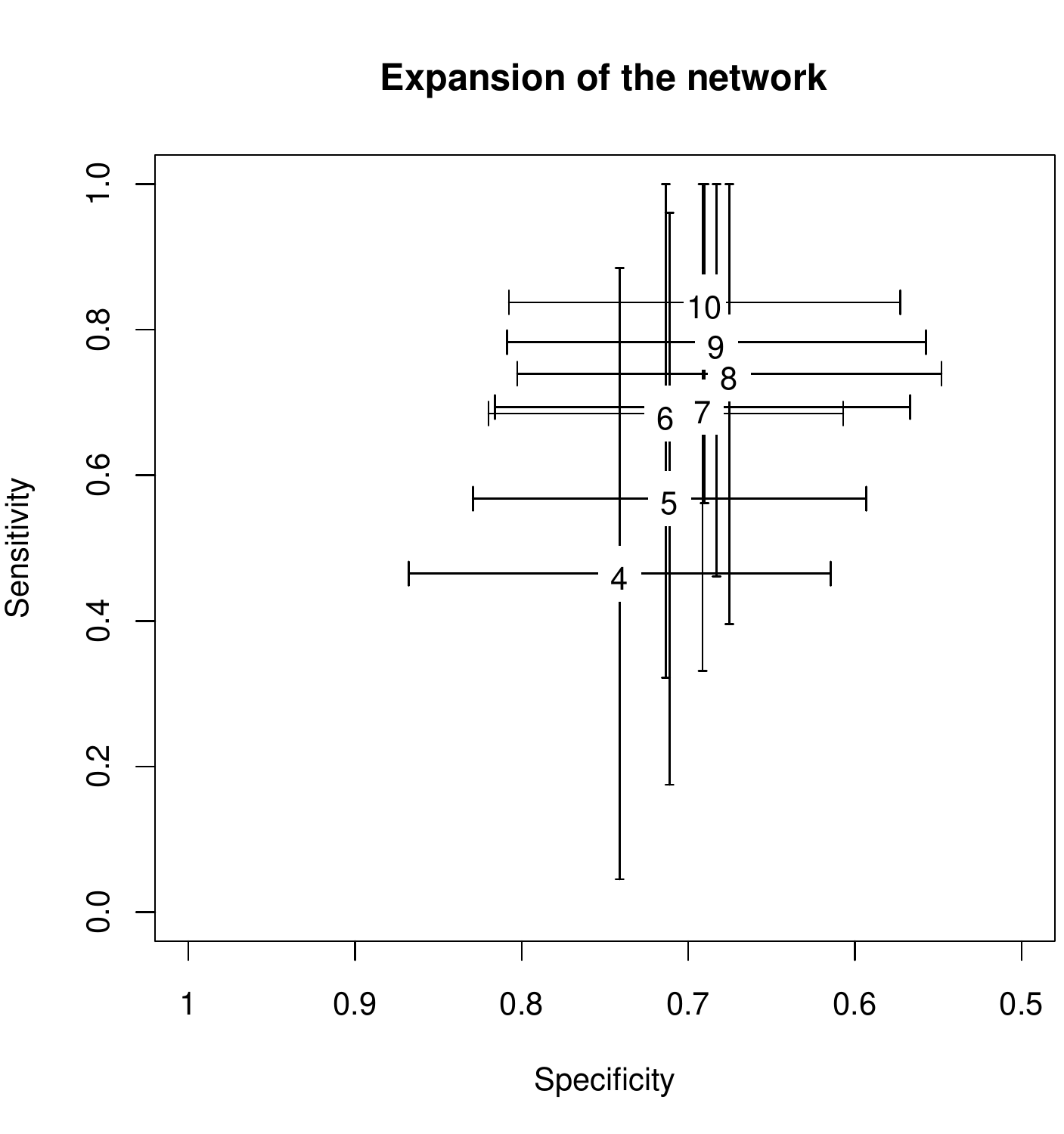}
\end{center}
\caption{Effects of prior knowledge. \textbf{Left:} Keep a fixed
model, and increase the number of known present edges ($0,...,4$).
\textbf{Middle:} Keep a fixed model, and increase the number of
known absent edges ($0,...,6$). \textbf{Right:} Start with an
unknown graph of $4$ actions, and add new actions ($0,...,6$) as
well as their adjacent edges to the graph. In all three plots, the
error bars range from the first to the third quartile of the
distributions obtained in 100 simulation runs.} \label{sensspec}
\end{figure}

The left and middle plot of Fig.~\ref{sensspec} show the results of
this procedure. The left plot illustrates the reconstruction quality
in dependence of the number of a priori known present edges. The
middle plot does the same for the inclusion of prior knowledge about
absent edges. Both plots show that including prior information
considerably increases sensitivity and specificity. In particular,
information about present edges helps more than information about
missing edges.

In a second experiment, a fixed ``true" model of $10$ actions and
$15$ edges is created, and a noise ratio matrix is generated from
it. We randomly pick a subgraph of $4$ actions, the structure of
which is assumed to be completely unknown. Another $k$ nodes
($k=0,...,6$) are added to the subgraph, and all edges not belonging
to the initial subgraph are correctly specified as known
present/absent via the actions graph prior. For each $k$, we
restrict the original ratio matrix to the nodes present in the
$(k+4)$-nodes subgraph and start an exhaustive search.

Again, the quality of reconstruction is reported by sensitivity and
specificity averaged over 100 sample runs. The noise level was set
to $0.4$, and the number of observables was set to $200$. The
results in the right plot of Fig.~\ref{sensspec} show a strong
increase in sensitivity at the cost of a slight decrease in
specificity.

\section{Application to {\em Drosophila} immune response}\label{section:application}

We apply our methodology to data from an RNA interference (RNAi)
gene silencing study on innate immune response in \emph{Drosophila
melanogaster} \citep{boutros02sequential}. The experiment probes how
transcriptional response to lipopolysaccharides (LPS) is regulated
by signal transduction pathways in the cell.

\paragraph{Data}
The data set consists of 16 Affymetrix microarrays: 4 replicates of
control experiments without LPS and without RNAi (negative
controls), 4 replicates of expression profiling after stimulation
with LPS but without RNAi (positive controls), and 2 replicates each
of expression profiling after applying LPS and silencing one of the
four candidate genes \emph{tak}, \emph{key}, \emph{rel}, and
\emph{mkk4/hep}.

Selectively removing one of these signaling components blocks
induction of all, or only parts, of the transcriptional response to
LPS. \cite{boutros02sequential} show that  this observation can be
explained by a fork in a signaling pathway below \emph{tak}, with
\emph{key} and \emph{rel} on the one side and \emph{mkk4/hep} on the
other. This result clarified the contributions of different pathways
to immune response in \emph{Drosophila} \citep{royet05sensing}.

\paragraph{Previous analyses}
The experimental design of this study, which includes both negative
and postive controls, allows to define informative effects of
interventions and quantify the false positive and false negative
rates. In the original analysis \citep{boutros02sequential} and two
subsequent studies
\citep{markowetz05nontranscriptional,markowetz07nested} only the 68
genes differentially expressed between positive and negative
controls were used as effect reporters.
\cite{markowetz05nontranscriptional} propose a simple discretization
scheme based on the two controls: if by silencing a gene in the LPS
stimulated cell the expression of an LPS-inducible gene moved close
to its expression in the negative controls, this was counted as an
\emph{effect} of the intervention; if a gene's expression stayed
close to its expression in the positive controls, the gene was
counted as being \emph{not affected} by the intervention. Applying
the same discretization scheme to the positive and negative controls
makes it possible to estimate the two error rates.

\paragraph{Analysis based on a single control}
The two types of controls can be used to define a set of informative
effect reporters and assess the error rates in the data. However,
most experimental studies do not contain two kinds of controls but
only one. To mimic this situation we will make no use of the
negative controls in the dataset and only include the four
LPS-induced measurements in our analysis. We show in the following
that our improved methodology is still applicable and exploits the
information in the data better  than previous approaches.

\begin{figure}[t]
\centerline{\includegraphics[width=.8\textwidth]{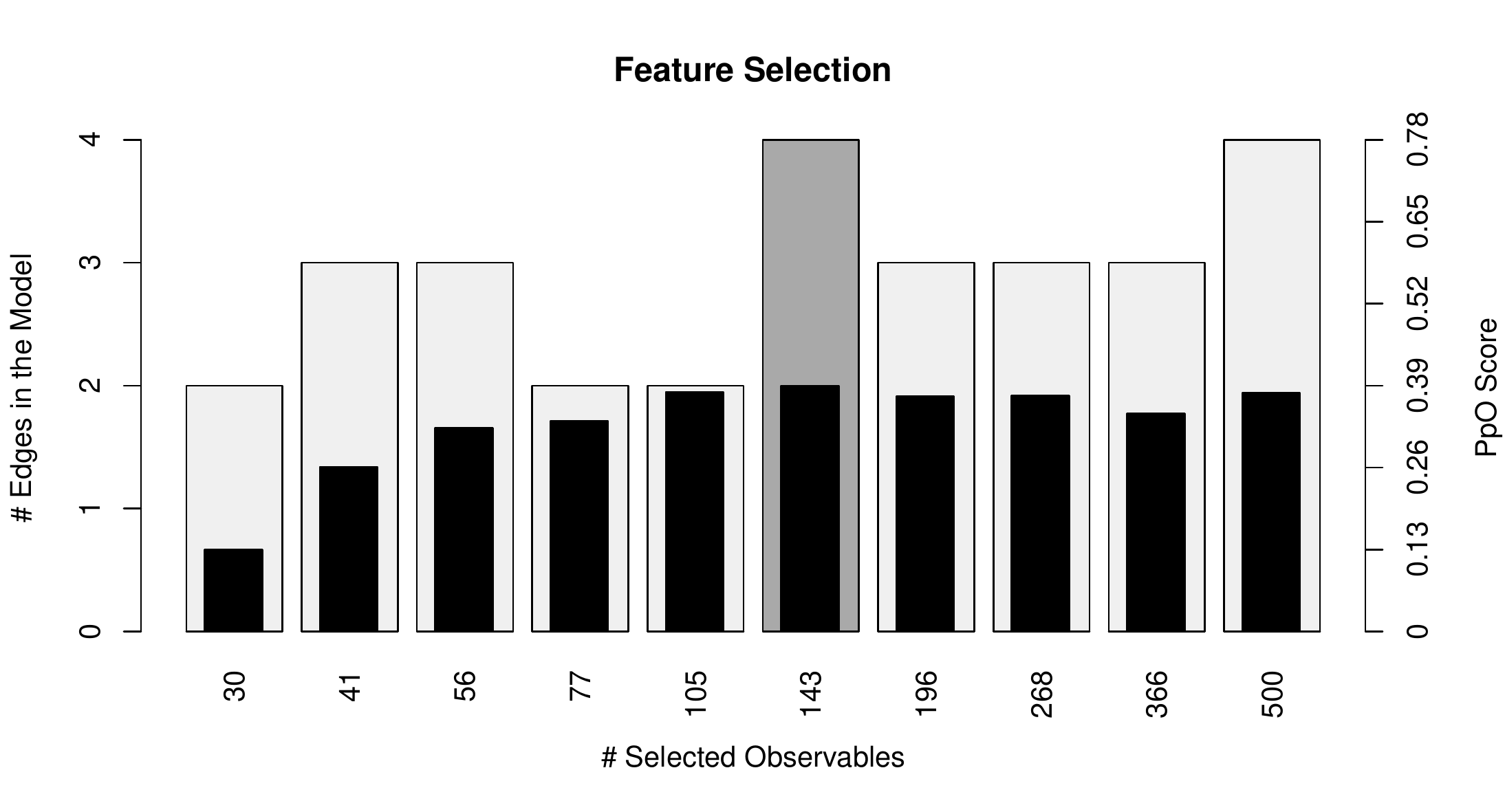}}
\caption{\label{deltachoice} Application of automatic regularization
to the \emph{Drosophila} data set. Each column corresponds to one
value of $\delta$ and a selected number of observables between $30$
and $500$. For each value of $\delta$ we plot (1.) the $\ppo$ score
(black bar with scale on the right) and (2.) the number of edges in
the inferred model (gray bar with scale on the left). The dark grey
bar at $143$ observables indicates the optimal degree of
regularization. The corresponding model is discussed in
Fig.~\ref{fig:boutrosresult}. }
\end{figure}

\begin{figure}[t]
\centerline{\includegraphics[width=.7\textwidth,height=0.4\textheight]{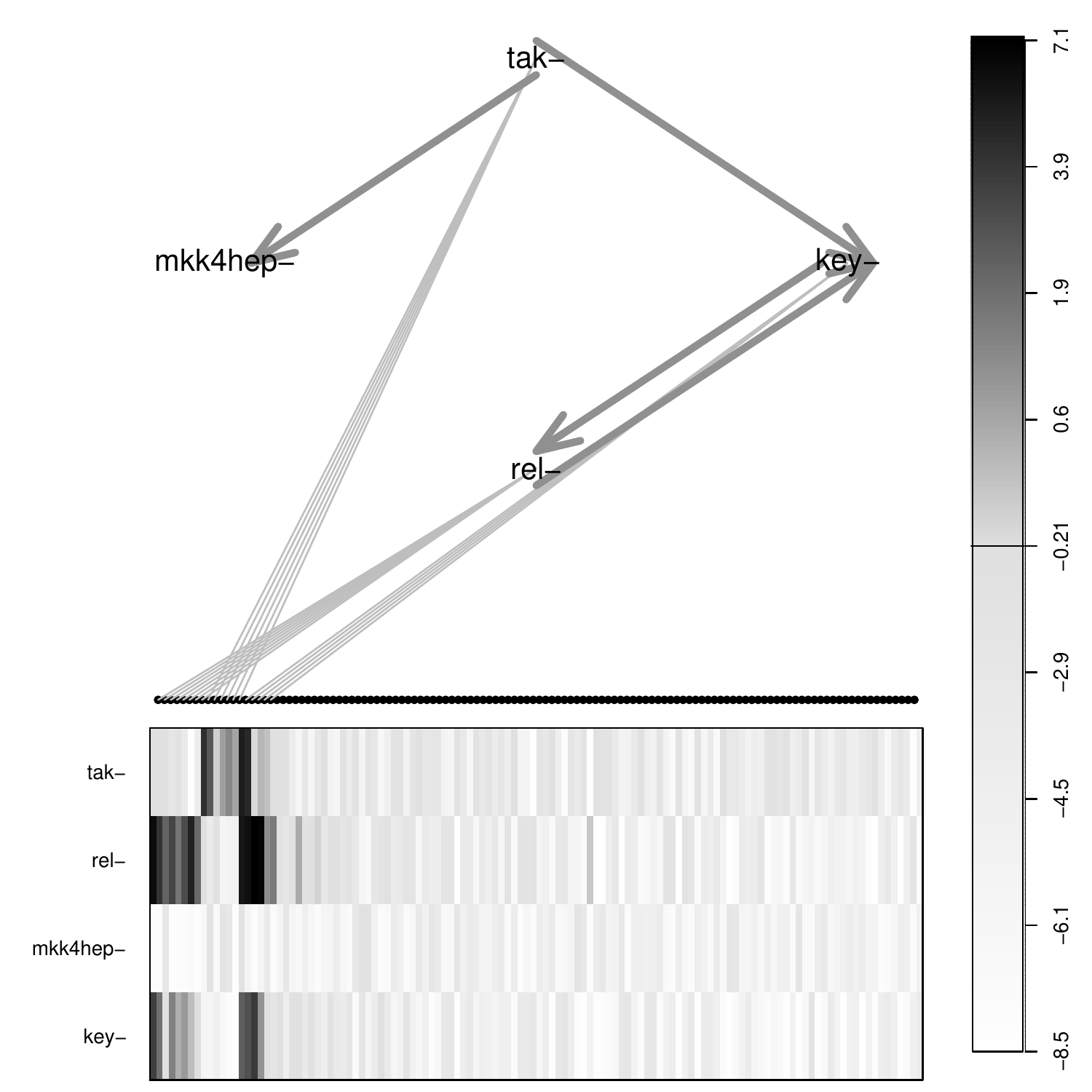}}
\caption{\label{fig:boutrosresult} Results on \emph{Drosophila}
data. The upper graph represents $\hat{\Gamma}$ on $\act =
\{\text{key-}, \text{tak-}, \text{rel-}, \text{mkk4/hep-} \}$, while
the assignment $\hat{\Theta}$ is shown as grey lines connecting
nodes in $\hat{\Gamma}$ with observables. The matrix below shows the
ratio matrix $R$ (where each column is one observable) with darker
values of grey indicating higher likelihood rations (see the
colorbar on the right). The graph $\hat{\Gamma}$ places \emph{tak}
above all other nodes and shows a branch below \emph{tak} with
\emph{key} and \emph{rel} on one side and \emph{mkk4/hep} on the
other side. The double headed arrow between \emph{key} and
\emph{rel} shows that the model can not distinguish between  them
(see the nearly identical rows in the ratio matrix). }
\end{figure}

\paragraph{Calculation of the ratio matrix $R$}\label{rmatrix}
We use well established methods to assess differential gene
expression between the positive controls (LPS stimulation but no
gene silencing) and the gene perturbation profiles. Because of the
small number of samples we chose a highly regularized empirical
Bayes method for assessing differential expression in microarray
experiments \citep{smyth04linear}, which is implemented in the
R-package \texttt{limma} \citep{smyth05limma} available from
\texttt{www.bioconductor.org}. The empirical Bayes approach is
equivalent to shrinkage of the estimated sample variances towards a
pooled estimate, resulting in far more stable inference when the
number of arrays is small. We compute likelihood ratios for the
comparison of positive controls against every gene perturbation. We
then select genes which show a positive ratio (regardless of its
size) for at least two of the four knock-downs. This simple step of
deleting uninformative genes reduces the number of effect reporters
(observables) from 14\ 010 to 904. This number is still much bigger
than the number of differential genes used in previous analyses and
makes feature selection necessary.

\paragraph{Results} We fit NEM models to the ratio matrix $R$
using the feature selection mechanism described in
section~\ref{section:featureselection}. The resulting curve of the
$\ppo$ statistic is shown in Fig~\ref{deltachoice}. The model
selected in our automatic procedure includes 143 observables (out of
904) and is shown in Fig.~\ref{fig:boutrosresult}.

Our model places \emph{tak} above all other nodes and shows a branch
below \emph{tak} with \emph{key} and \emph{rel} on one side and
\emph{mkk4/hep} on the other side. The gene perturbations \emph{key}
and \emph{rel} remain undistinguishable due to almost identical
phenotypic profiles (see the nearly identical rows in the ratio
matrix in Fig.~\ref{fig:boutrosresult}). The branching below
\emph{tak} into two sub-pathways is the main biological feature of
the data \citep{boutros02sequential} and our model succeeds in
recapitulating it.

A previous analysis of the same data set
\citep{markowetz05nontranscriptional} showed a very similar picture
but included one additional edge from \emph{tak} to \emph{rel},
which is not contained in our result. It is known that \emph{rel} is
a transcription factor responsible for immune response, which is
activated via the kinase \emph{key} \citep{royet05sensing}. Thus,
the additional edge from \emph{tak} to \emph{rel} was a spurious
result. It can be explained by the fact that the NEMs used in
\citep{markowetz05nontranscriptional} were constrained to
transitively closed graphs (and then the direct edge from \emph{tak}
to \emph{rel} is needed because there is a path from \emph{tak} to
\emph{rel} over \emph{key}). This shows that our general formulation
of NEMs, which is not constrained to transitively closed graphs, can
yield results closer to biological reality than previous
formulations.

\section{Discussion}

In this paper we introduced a generalized definition of Nested
Effects Models and expanded  their statistical basis in several
important directions. Our most important theoretical result is that
NEMs can be shown to be identifiable under mild conditions on the
data.

\paragraph{General NEMs}
The new general formulation of NEMs expands the model class from
transitively closed graphs to all directed graphs. This reduces the
bias in the model and leads to results closer to existing biological
knowledge in the application to {\em Drosophila} immune response.

\paragraph{Likelihood formulation}
The new likelihood equation is much more flexible than previous
equations for binary data
\citep{markowetz05nontranscriptional,markowetz07nested}. It is
applicable to any kind of data by converting it into likelihood
ratios for the comparison of effects and non-effects. Thus, it can
even integrate heterogeneous sources of data as long as they can be
translated into likelihood ratios. Additionally, missing data or the
exclusion of bad measurements is possible without changes in the
algorithm.

\paragraph{Model search}
Our formulation of the likelihood also leads to a fast updating
procedure which can be carried out in linear time and is exceedingly
faster than previous approaches. Still, an exhaustive search is
clearly infeasible for larger values of perturbed genes. However,
the fast elementary moves introduced here allow the application of
combinatorial search algorithms, like Markov Chain Monte Carlo
\citep{gilks96mcmc} or simulated annealing
\citep{kirkpatrick83optimization}, to find high scoring models.

\paragraph{Prior knowledge}
We showed the usefulness of incorporating prior knowledge into
model search by fixing parts of a bigger model and only inferring
the unknown part. This is a special case of a prior distribution on
the space of model graphs. We hope to extend this approach to more
flexible structure priors. One promising research direction could be
to use a structure prior that favors transitively closed graphs. In
this way it would be possible to find a balance between the less
biased models introduced here and the causally interpretable but
more constrained models introduced earlier
\citep{markowetz05nontranscriptional,markowetz07nested}.

\paragraph{Availability of software}
The NEM exhaustive search algorithm and all its extensions described
in this paper are implemented, documented and ready to use in the R
package \texttt{Nessy}, which is available at
\texttt{www.bioconductor.org}. It includes a plotting routine that
conveniently displays nested effects models (see, e.g.,
Fig~\ref{fig:boutrosresult}).

\section*{Appendix: Proofs of Theorems}

\paragraph{Theorem~\ref{consistency}}

{\em If the data is consistent with the effects model $F$, then the
maximum likelihood estimate of (\ref{trace1}) equals $F$,
\[
 F =\ \underset{G}{\argmax}\ P(D|G)\ \underset{(\ref{trace1})}{=}\  \underset{G}{\argmax}\ \tr(GR)\ .
 \]
}

\begin{proof}
We have $\tr (GR) =  \sum_{a\in\act}\sum_{s\in\,\obs} G_{a s}
R_{sa}$.

If $F_{as}=1$, then by consistency of the data $R_{sa}>0$ and the
choice $G_{as}=1=F_{as}$ maximizes the summand $G_{as} R_{sa}$.

If $F_{as}=0$, then $R_{sa}\leq 0$ and the coice $G_{as}=0=F_{as}$
maximizes the summand $G_{as} R_{sa}$. Hence $\underset{G}{\argmax}\
\tr(GR) = F$.
\end{proof}

\paragraph{Lemma~\ref{lemma:reversal}} {\em Let $(\Gamma',\Theta')$ be a reversal of $(\Gamma,\Theta)$ induced by the permutation $\pi = (a_1,a_2,...,a_n)$. Then $(\Gamma',\Theta')$ is a valid parametrization
of $F=\Gamma\Theta$ }.

\begin{proof}
Let $e_a$ denote the $a$-th unit column vector of length $n_{\obs}$.
Clearly, $\Gamma'\Theta' = (\Gamma S^{-1})(S\Theta) = \Gamma \Theta
= F$. The only additional requirement we need to check is
$\Gamma'_{aa}=1$ for all $a\in \act$. This holds because of
\begin{eqnarray*}
\Gamma'_{aa} & = & e_a(\Gamma S^{-1})e_a^T  \; = \; e_a\Gamma
\sum_{b\in \act} e_{\pi(b)}^T e_b e_a^T \; = \; e_a\Gamma
e_{\pi(a)}^T \; = \;
\\ &=& \Gamma_{a \pi(a)}\; = \; \begin{cases} \Gamma_{a_j \pi(a_j)} &
\text{if }a\in\{a_1,...,a_n\} \\ \Gamma_{aa} &
\text{otherwise}\end{cases}\ \;  = \; 1.
\end{eqnarray*}
\end{proof}

\paragraph{Theorem~\ref{identifiability}}
{\em Let $(\Gamma,\Theta)$ and $(\Gamma',\Theta')$ be the parameters
of two nested effects models. Assume that no two distinct actions
$a,b\in\act$ have the same parents in $\Gamma$ or in $\Gamma'$. Then
$(\Gamma,\Theta)$ and $(\Gamma',\Theta')$ are observationally
equivalent if and only if the tuples can be converted one into
another by a sequence of disjoint reversals.}

\begin{proof}
"$\Leftarrow$": This follows immediately from Lemma~\ref{lemma:reversal}.\\
"$\Rightarrow$": For an action $a\in \act$, denote the parents of $a$ in $\Gamma$ by
$\parents_{\Gamma}(a)=\{b\in\act\fdg b\stackrel{\Gamma}{\to}a \}$.
Recall that for an actions graph $\Gamma$,
$a\in\parents_{\Gamma}(a)$ for all $a\in\act$. The set of
observables attached to $a$ via the effects graph $\Theta$ is called
the children of $a$ in $\Theta$, $\children_{\Theta}(a) =
\{s\in\obs\fdg a\stackrel{\Theta}{\to} s\}$. Since
$\parents_{\Gamma}(a)$ determines $\Gamma_{\cdot a}$ (and vice
versa), and $\children_{\Theta}(a)$ determines $\Theta_{a\cdot}$
(and vice versa), the family of all parents sets determines $\Gamma$
and the family of all children sets determines $\Theta$.\\
Assume that $\children_{\Theta}(a)$ and
$\children_{\Theta'}(b)$ intersect nontrivially, say
$a\stackrel{\Theta}{\to} s$, $b\stackrel{\Theta'}{\to} s$. Then
\begin{eqnarray}
\Gamma_{\cdot a} = \Gamma_{\cdot \theta(s)} = (\Gamma\Theta)_{\cdot
s} = (\Gamma'\Theta')_{\cdot s} = \Gamma'_{\cdot \theta'(s)} =
\Gamma'_{\cdot b}
\end{eqnarray}
Hence
\begin{eqnarray}\label{thm.1}
\parents_{\Gamma}(a) = \parents_{\Gamma'}(b)
\end{eqnarray}
Furthermore, let $t\in \children_{\Theta}(a)$ and $t\in \children_{\Theta'}(c)$.
Then by (\ref{thm.1}), $\parents_{\Gamma}(a) =
\parents_{\Gamma'}(c)$, which in turn together with (\ref{thm.1})
implies $\parents_{\Gamma'}(b) = \parents_{\Gamma'}(c)$. By the
hypothesis, this is only possible if $b=c$. Thus
$\children_{\Theta}(a)\subseteq \children_{\Theta'}(b)$, and for
symmetric reasons, $\children_{\Theta}(a) = \children_{\Theta'}(b)$.
It follows that the partitions
\begin{eqnarray}
  \cdju{a\in\act}{} \children_{\Theta}(a) = \obs =  \cdju{a\in\act}{} \children_{\Theta'}(a)
\end{eqnarray}
are identical up to order. Therefore there exists a permutation
$\pi$ of \act such that
\begin{eqnarray}\label{thm.2}
    \children_{\Theta'}(a) = \children_{\Theta}(\pi(a))\ \ ,\ a\in\act
\end{eqnarray}
Together with (\ref{thm.1}) this implies
\begin{eqnarray}\label{thm.3}
\parents_{\Gamma'}(a) = \parents_{\Gamma}(\pi(a))\ \ ,\ a\in\act
\end{eqnarray}
In other words, $(\Gamma',\Theta')$ is determined completely by
$(\Gamma,\Theta)$ and $\pi$. Let us investigate $\pi$ a little
closer. Let $T = \sum_{b\in\act} e_be_{\pi(b)}^T$ be the permutation
matrix associated with $\pi$. We confirm that
\begin{eqnarray}
(T\Theta)_{a\cdot} = e_a^T \sum_{b\in\act} e_be_{\pi(b)}^T \Theta =
e_{\pi(a)}^T \Theta = \Theta_{\pi(a)\cdot}
\stackrel{(\ref{thm.2})}{=} \Theta'_{a\cdot}\ \ ,\ a\in\act\ ,
\end{eqnarray}
so $\Theta' = T\Theta$. By analogous calculations, $\Gamma' = \Gamma
T^{-1}$. Now for $a\in\act$,
\begin{eqnarray}
    1 = \Gamma'_{aa} \stackrel{(\ref{thm.3})}{=} \Gamma_{a \pi(a)}\ ,
\end{eqnarray}
which means that there exists an edge from $a$ to $\pi(a)$ in
$\Gamma$. For any $a\in\act$, $\Gamma$ must therefore contain the
cycle $a\to\pi(a)\to \pi^2(a)\to ... \to a$.

Let $\pi = \pi_1\pi_2\cdot...\cdot \pi_m$ be a decomposition of
$\pi$ into disjoint cycles, and let $T_1,T_2,...,T_m$ be the
$n_{\act}\times n_{\act}$ permutation matrices corresponding to
$\pi_1,\pi_2,...,\pi_m$ respectively. Clearly,
$T=T_1T_2\cdot...\cdot T_m$.

Define $(\Gamma_0,\Theta_0) = (\Gamma,\Theta)$, and inductively
$(\Gamma_j,\Theta_j) = (\Gamma_{j-1}T_j^{-1},T_j\Theta_{j-1})$,
$j=1,...,m$. Then $(\Gamma_m,\Theta_m)=(\Gamma
T^{-1},T\Theta)=(\Gamma',\Theta')$, and we have constructed a
sequence of disjoint reversals converting $(\Gamma,\Theta)$ into
$(\Gamma',\Theta')$.
\end{proof}

\section*{Acknowledgements}

AT would like to thank Olga Troyanskaya's lab in Princeton for the
excellent hospitality during the preparation of this paper. Both
authors greatly appreciated the discussions with all members of the
group, in particular Maria Chikina, Edo Airoldi, Patrick Bradley and
Chad Myers.

FM is supported by  NIH grant R01 GM071966 and NSF grant IIS-0513552
to O. G. Troyanskaya (Lewis-Sigler Insitute for Integrative Genomics
and Dept. of Computer Science, Princeton University, Princeton, NJ
08544, USA). This research was partly supported by NIGMS Center of
Excellence grant P50 GM071508 and by NSF grant DBI-0546275.


\end{document}